\documentclass[11pt,a4paper]{article}
\usepackage[utf8]{inputenc}
\usepackage[a4paper, total={7in, 10in}]{geometry}
\usepackage{mathtools, amsmath, amssymb, hyperref, graphicx }
\usepackage{subfig}
\usepackage{cite}
\usepackage[T1]{fontenc}
\usepackage{authblk}
\hypersetup{
                    colorlinks=true,
                    linkcolor=blue,
                    urlcolor=cyan
                    }

\title{\bf{Effect of Minimal Length on Landau Diamagnetism and de Haas-van Alphen Effect }}

\author[a,b,c]{Md. Abhishek\thanks{mdabhishek@hri.res.in}}
\author[a]{Bhabani Prasad Mandal\thanks{bhabani.mandal@gmail.com}}
\affil[a]{\small{\em Department of Physics, Institute of Science, Banaras Hindu University, Varanasi, India-221005.}}
\affil[b]{\small{\em Harish-Chandra Research Institute, Chhatnag Road, Jhunsi, Allahabad, India-211019.}}
\affil[c]{\small{\em Homi Bhabha National Institute, Training School Complex, Anushaktinagar,
Mumbai, India-400094.}}

\begin{document}

\maketitle

\begin{abstract}
	
We study Landau diamagnetism in the framework of generalised uncertainty principle(GUP). We calculate the correction to magnetisation and susceptibility by constructing  the grand partition function of diamagnetic material in this framework. We explicitly  show that Curie's law gets a temperature independent correction which vanishes when quantum gravity effects are neglected. We further consider the low temperature limit to find  how GUP affects the de Haas-van Alphen effect.

\end{abstract}


\section{Introduction}

All the basic forces except gravity are well described in the quantum framework whereas the theory of gravity was understood based on Einstein's General theory of relativity which is formulated in classical physics. There were certain areas where we need to merge quantum mechanics and general relativity to develop the quantum theory of gravitation. The existence of a minimum possible length, the Planck length $l_{pl}\approx 10^{-35}m$ \cite{AMATI198941,Gross:1987kza,Gross:1987ar,Tawfik_2014,Konishi:1989wk,Ali_2009,PhysRevD.84.044013}, was predicted by all approaches of quantum gravity theory, doubly special relativity(DSR), perturbative string theory, black hole physics, etc. The usual Heisenberg's uncertainty principle(HUP) needs essential modifications to incorporate the existence of minimum length scale. The new uncertainty principle, known as the generalised uncertainty principle(GUP) \cite{AMATI198941,AMELINO_CAMELIA_2002,GARAY_1995,Maggiore_1993,Scardigli_1999}, is as follows, 
\begin{eqnarray}
\Delta x\Delta p &\geq& \frac{1 }{2}[1+\alpha(\Delta p)^{2}+\alpha \langle p{\rangle}^{2}],\label{gup}
\end{eqnarray}
where $\Delta x $ and $ \Delta p $ are the uncertainties in position and momentum respectively. The positive parameter $\alpha = \frac{\alpha_{0}l_{pl}^{2}}{2} = \frac{\alpha_{0}}{M_{pl}^{2}},$ does not depend on $\Delta x$ and $\Delta p$ but it may depend on the expectation values of $\bf{x}$ and $\bf{p}$. $\alpha_{0}$ is known as the GUP parameter which is a positive dimensionless parameter, and the Planck mass $ M_{pl} \approx 10^{19} Gev$. According to Eq.(\ref{gup}), a nonzero minimal uncertainty in length is $\Delta x_{min}\approx l_{pl}\sqrt {\alpha_{0}}$. The parameter $\alpha_{0}$ is considered to be $1$ in most of the GUP calculations. There are several ways to construct the modified Heisenberg algebra due to GUP and the most commonly used  modified Heisenberg algebra due to GUP approach, via the Jacobi identity \cite{Scardigli_1999,PhysRevD.52.1108}  is given by,
\begin{align}\label{gupcom}
[x_{i},p_{j}] &= i (\delta _{ij}+\alpha \delta _{ij}p^{2}+2\alpha p_{i} p_{j}),\quad [x_{i},x_{j}] = 0, \quad [p_{i},p_{j}] = 0.
\end{align}
To be consistent with Eq.(\ref{gup}) we can define the position and the momentum operators in the following way \cite{PhysRevD.84.044013}\cite{PhysRevD.52.1108}\cite{PhysRevLett.101.221301},
\begin{align}\label{gupmom}
x_{i} &\equiv x_{0i},\quad p_{i} \equiv p_{0i}(1+\alpha p_{0}^{2}),
\end{align}
where, $p_{0}^{2} = \sum_{j=1}^{3} p_{0j}p_{0j}$. The operators $x_{0i}$ and $p_{0i}(= -id/dx_{0i})$, at low energy, obey the commutation relation, 
\begin{eqnarray}
[x_{0i},p_{0j}] &=& i\delta _{ij}.
\end{eqnarray}
In the presence of potential $V(\vec{r})$ the Hamiltonian of any quantum mechanical system should be modified due to GUP in the following form \cite{PhysRevLett.101.221301,Tawfik_2015,GECIM2017391,SCARDIGLI2017242,majumder2011effects,_vg_n_2017,MAZIASHVILI2006232,FAIZAL2017238,PhysRevD.87.065017,Pedram_2012,Faizal_2015,Bhat_2017,Bagchi_2009,Ghosh_2012,PhysRevD.91.124017,PhysRevD.87.084033,PhysRevD.86.064038,Chen_2014,NOZARI_2005,Das:2009hs,PhysRevD.96.066008,Magueijo_2004,Tyagi:2019wjj,Verma:2018bqh},
\begin{eqnarray}\label{gupham}
H &=& \frac{p_{0}^{2}}{2m} + V(\vec{r}) + \frac{\alpha}{m} p_{0}^{4} + \mathcal{O}(\alpha^{2}).
\end{eqnarray}
During the past several years GUP and its rich consequences have been studied extensively in almost all branches of physics, including Black hole physics \cite{Scardigli_1999,GECIM2017391,_vg_n_2017, MAZIASHVILI2006232,Chen_2014}, Cosmology \cite{majumder2011effects}, Relativistic quantum mechanics \cite{PhysRevD.87.065017,Verma:2018bqh,NOZARI_2005}, Non-Hermitian theories \cite{Faizal_2015,Bagchi_2009}, Squeezed and Coherent states \cite{Ghosh_2012,PhysRevD.91.124017,PhysRevD.87.084033,PhysRevD.86.064038}, Statistical mechanics \cite{Nouicer_2007,El-Nabulsi:2020zyh,Hamil:2020ldl,Hamil:2022jkc}. In quantum optics, experiments with a noticeable quantum gravity effect were proposed \cite{Pikovski_2012,Bosso_2017}. Higher-order GUP and its effect have been considered in \cite{Pedram_2012,Hamil:2020ldl,Hamil:2022jkc}. In the region of minimum length, exact solutions to several relativistic and non-relativistic problems along with a review of the subject have been obtained \cite{Tawfik_2015}.  The effect of minimum length on electron magnetism using the ideas of quantum statistical mechanics was considered in \cite{Nouicer_2007} to remove the degeneracy of Landau levels. More recently in \cite{Hamil:2022jkc}, the thermodynamics of ideal gas, Unruh-Davies-Dewitt Fulling effect, blackbody radiation has been investigated using the statistical mechanical formulation with GUP.  Various GUP-modified thermodynamical quantities have been compared using  three dissimilar  forms of GUP in the formulation of statistical mechanics \cite{El-Nabulsi:2020zyh, bouaziz}.  New higher-order GUP formulation in statistical mechanics is considered to calculate the canonical partition function  for an ideal gas in one dimension and hence GUP modified thermodynamical quantities were calculated in  \cite{Hamil:2020ldl}. In this article, we revisit the effect of minimum length on diamagnetism  by constructing the grand partition function and studying Curie's law and the de Haas-van Alphen effect. 

 In this paper, we take the velocity of light in vacuum, $c=1$, reduced Planck constant, $\hbar = 1$, and Boltzmann constant, $k_B=1$ throughout our calculation. In section \ref{diamag}, we have calculated the grand-canonical partition function of a diamagnetic material and the corrections to the specific heat, magnetisation and susceptibility due to minimal length at high temperature. The effect of GUP on the low temperature  de Haas-van Alphen effect has been studied in section \ref{dhv effect}. We end with concluding remarks in section \ref{conclusions} where we summarise our main results and discuss some future directions. 


\section{Diamagnetic properties at high temperatures}\label{diamag}

 In this section, we study the diamagnetic property of material under the GUP effect. For this, we first calculate the grand-canonical partition function of a system of $N$ spinless electrons with mass $m$, confined in a volume $V$ under the action of a constant magnetic field  $B$ along the $z$-direction. Using the grand-canonical partition function we systematically derive the susceptibility $\chi$ at a high temperature limit. When we turn on the magnetic field the electrons of the system start to move on a helical path with its axis parallel to the direction of the magnetic field. The projection of the helical trajectory of the electrons to the $(x,y)$ plane becomes circular\cite{huang2008statistical}\cite{greiner2012thermodynamics}. Now using the modified commutation relation due to quantum gravity, the single particle energy levels of electrons are given by\cite{PhysRevLett.101.221301}\cite{Das:2009hs},
\begin{eqnarray}
\epsilon _{j} = \frac{p_{z}^{2}}{2m}+2\mu B(j+\frac{1}{2})+\alpha\bigg[\frac{p_{z}^{4}}{m}+16m\mu^{2}B^{2}(j+\frac{1}{2})^{2}\bigg ] + \mathcal{O}(\alpha^{2}).
\end{eqnarray}
These quantised energy levels with quantum number $j$ are degenerate. Now the density of the one particle states \cite{ali2011minimal}, 
\begin{align}
g_{j} & =  \frac{1}{4\pi^2}\int_{p_{j}}^{p_{j+1}} \frac{dp_{x}dp_{y}dxdy}{(1+\alpha p^{2})^{3}}\nonumber\\
&= \frac{ V^{2/3}}{\pi}m\mu B[1 - 6\alpha m\mu B(2j+1)]+ \mathcal{O}(\alpha^{2}),
\end{align}
where $p_{j} \equiv \sqrt {4m\mu B j}\ $ and the Bohr magneton, $\ \mu\equiv\frac{e}{2m}$. As opposed to the low energy case, we can see that $g_{j}$ is dependent on the quantum number $j$. At finite temperature $T$ the grand-canonical partition function of the electron gas becomes, 
\begin{eqnarray}
\ln \mathcal{Z}_{G} = \int^{+\infty}_{-\infty}\frac{V^{1/3}dp_{z}}{2\pi(1+\alpha p_{z}^{2})^2}\sum_{j=0}^{\infty}g_{j}\ln[1+\mathtt{z}\ \exp(-\beta \epsilon _{j})]
\end{eqnarray}
At a high temperature or equivalently at low $\beta(=\frac{1}{T})$, we can take the fugacity, $ \mathtt{z}<<1 $, and the system is effectively Boltzmannian, thus,
\begin{align}\label{partition}
\ln \mathcal{Z}_{G} = & \int^{+\infty}_{-\infty}\frac{\mathtt{z}\ V^{1/3}dp_{z}}{2\pi(1+\alpha p_{z}^{2})^2}\sum_{j=0}^{\infty}g_{j}\  \exp(-\beta \epsilon _{j})\nonumber\\
= &\frac{V \mathtt{z}}{\lambda ^3}x\ \text{cosech}\ x\bigg[1 - \frac{m\alpha}{\beta}\{5 + 6x\coth{x} + 4x^{2}(\coth^{2}{x} + \text{cosech}^{2}{x})\}\bigg].
\end{align}
To make the above expression compact we have introced the thermal wavelength  $\ \lambda \equiv \sqrt{\frac{2\pi}{ {mT}}}$ and a new variable $\ x \equiv \frac{\mu B}{T}$. In case of $\alpha \rightarrow 0$, the above equation is reduced to the grand-canonical partition function of the diamagnetic material without GUP effect \cite{huang2008statistical,greiner2012thermodynamics}. Now the average number of electrons in the system is,
\begin{eqnarray}
&&\bar{N}  =  \bigg(\mathtt{z} \frac{\partial }{\partial \mathtt{z}}\ln \mathcal{Z}_{G}\bigg)_{T,V,B}.
\end{eqnarray}
At high temperature limit, fugacity  $\mathtt{z}\rightarrow 0$ and if the field intensity $B$ and the temperature $T$ are such that $\mu B<< T$ then,
\begin{eqnarray}
&&\bar{N} \simeq  \frac{V\mathtt{z}}{\lambda^3} {\big[}1 - 19\frac{\alpha}{\beta} m {\big]}.
\end{eqnarray}
To study thermodynamic properties of the system at high temperature, let us compute the internal energy of the system in terms of variable $x$,
\begin{align}
 U  =& -\bigg(\frac {\partial}{\partial \beta}\ln \mathcal{Z}_{G}\bigg)_{V,\mathtt{z}}\nonumber\\
  =&-\mu B\bigg(\frac {\partial}{\partial x}\ln \mathcal{Z}_{G}\bigg)_{V,\mathtt{z}}\nonumber\\
 =& V \mathtt{z}\left(\frac{m}{2\pi}\right)^{3/2}(\mu B)^{5/2}\left(\frac{ \text{cosech}\ x}{2x^{5/2}}\right)\bigg[\left(x+2 x^2\ \text{coth}\ x\right)-\alpha m\mu B\bigg\{15+8 x^2\ \text{coth}^2\ x+8 x^3\ \text{coth}^3\ x\nonumber\\
&\hspace{6cm}+8 x^2\ \text{cosech}^2\ x+8 x\ \text{coth}\ x \left(2+5 x^2\ \text{cosech}^2\ x\right)\bigg\}\bigg].
\end{align}
Now the specific heat at constant volume,
\begin{align}
 C_V  =& \bigg(\frac {\partial U}{\partial T}\bigg)_{V}\nonumber\\
 =&-\frac{\mu B}{T^2}\bigg(\frac {\partial U}{\partial x}\bigg)_{V}\nonumber\\
 =&\frac{\Bar{N}}{4}\bigg[\left(3 + 4x \ \text{coth}\ x + 4x^2 \ \text{coth}^2\ x + 4x^2 \ \text{cosech}^2\ x\right) \nonumber\\
  &\hspace{1cm}- \frac{4\alpha m\mu B}{x}\Big\{15 + 10 x \ \text{coth}\ x  - 8 x^3 \ \text{coth}^3\ x  + 2 x^2 \ \text{cosech}^2\ x  \nonumber\\
   &\hspace{3cm}+ 16 x^4 \ \text{cosech}^4\ x  + \text{coth}^2\ x  (-4 x^2 + 64 x^4\ \text{cosech}^2\ x ) \Big\}\bigg].
\end{align}
The GUP corrected specific heat $(C_V)$ differs from without GUP value at very high temperatures in a constant magnetic field. But without the quantum gravity effect, $C_V$ acquires a constant value of $\frac{15}{4}\bar{N}$ as follows,
\begin{figure}
\centering
\subfloat[\;]{
  \includegraphics[scale=0.9]{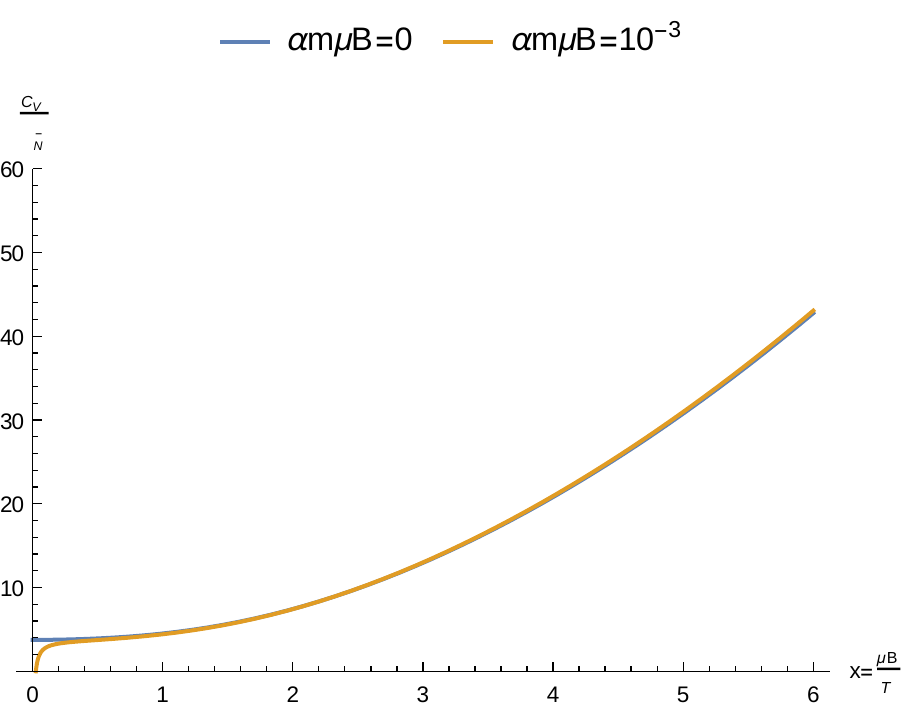}\label{a}  
}
\subfloat[\;]{
 \includegraphics[scale=0.9]{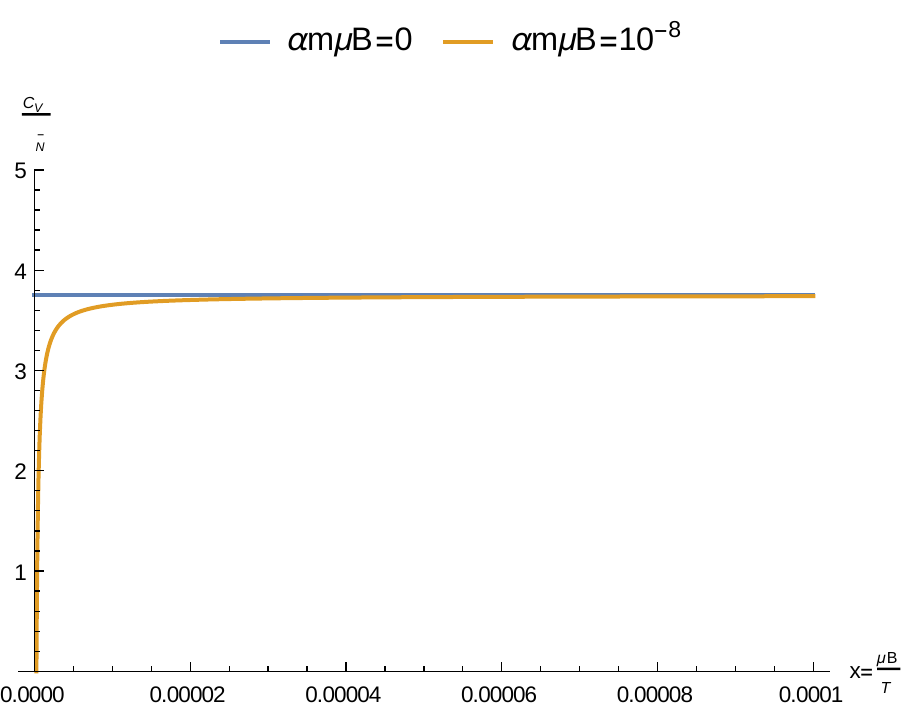}\label{b}
}
\caption{Specific heat at constant volume, $C_V/{\Bar{N}}$ vs $x=\frac{\mu B}{T}$ plot of a diamagnetic system with electron mass, $m=0.5\times10^{-3}Gev$ . In (a) we set the GUP dependent parameter, $\alpha mB \mu=10^{-3}$ and show the full $C_V/{\Bar{N}}$ vs $x=\frac{\mu B}{T}$ curve along with $\alpha=0$ curve. (b) shows the zoomed view nature of the two curves with $\alpha mB \mu=0$ and $\alpha mB \mu=10^{-8}$ at extremely  high temperature. }
\label{fig }
\end{figure}
\begin{align}
 C_V = & \frac{\bar{N}}{4}\left[3+4 x\ \text{coth}\ x+4 x^2\left( \text{coth}^2\ x+\ \text{cosech}^2\ x\right)\right]\nonumber\\
 \simeq & \frac{15}{4}\bar{N};\quad \text{for\ } \mu B<< T.
\end{align}
Let us now study the magnetic behaviour of the system in the presence of quantum gravity. Magnetic moment $M$ of the gas is given by,
\begin{eqnarray}
 M = \frac{1}{\beta}\bigg(\frac {\partial}{\partial B}\ln \mathcal{Z}_{G}\bigg)_{T,V,\mathtt{z}}.
\end{eqnarray}
For high temperature limit, the magnetic moment per unit volume, i.e the magnetisation is,
\begin{eqnarray}
 &\mathcal{M} = \frac{M}{V} = - \frac{1}{3}\frac{\mathtt{z}}{\lambda ^{3}} \mu x\left[ 1+ \frac{m\alpha}{\beta}\right].
\end{eqnarray}
Now at high temperature, the magnetic susceptibility per unit volume as a function of temperature $T$ and specific volume $\mathcal{V} = {V}/{\bar{N}}$ becomes,
\begin{align}\label{diasuscep}
\chi &= \bigg (\frac{\partial\mathcal{M}}{\partial B}\bigg )_{T,V,\mathtt{z}}\nonumber\\
&= - \frac{\mu^{2}}{3 \mathcal{V}  T}  \bigg[ 1 + 20\alpha mT \bigg].
\end{align}
The effect of GUP on this phenomenon is very tiny to detect experimentally in the lab. To demonstrate this small GUP corrections graphically we have taken the GUP parameter $\alpha=1Gev^{-2}$ in Figure \ref{fig 1}. The $1/T$ dependence in Figure \ref{fig 1} confirms that Curie's law still holds with the effect of minimum length. Also the Curie constant  $( \frac{\mu^{2}}{3 \mathcal{V} })$  reflects the quantization of the orbits. The negative sign in Eq.(\ref{diasuscep}) implies the property of diamagnetism $(\chi <0)$ is not affected by quantum gravity. In the case of $\alpha\rightarrow 0$, the above equation is reduced to the low energy susceptibility per unit volume of the electron gas.
\begin{figure}
 \centering
  \includegraphics[scale=0.95]{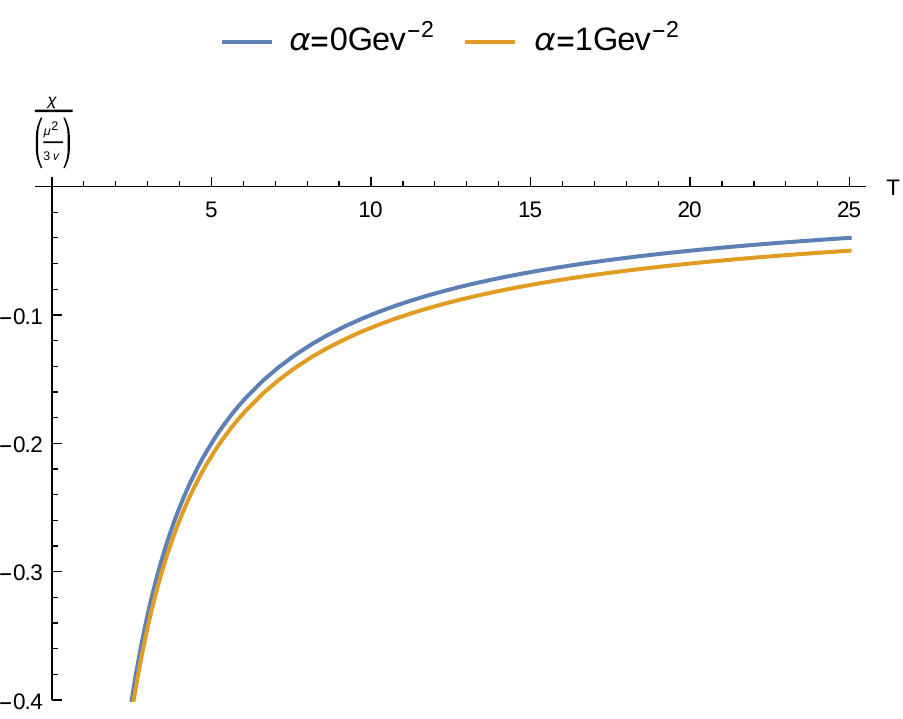}
  \caption{Reduced susceptibility, ${\chi}/{(\frac{\mu^{2}}{3 \mathcal{V}})}$ vs temperature, $T$ plot of Landau diamagnetism with electron mass, $m=0.5\times10^{-3}Gev$, and we set the GUP parameter, $\alpha=1Gev^{-2}$.}
  \label{fig 1}
\end{figure}
%

\section{Low temperature de Haas-van Alphen effect}\label{dhv effect}

At a very low temperature $(T\rightarrow 0)$ the susceptibility of an ideal Fermi gas  discontinuously changes with the varying magnetic field $B$, this phenomenon is known as the de Haas-van Alphen effect \cite{huang2008statistical}. In this section, we study the effect of GUP on this quantum mechanical effect.

\begin{figure}
 \centering
  \includegraphics[scale=0.8]{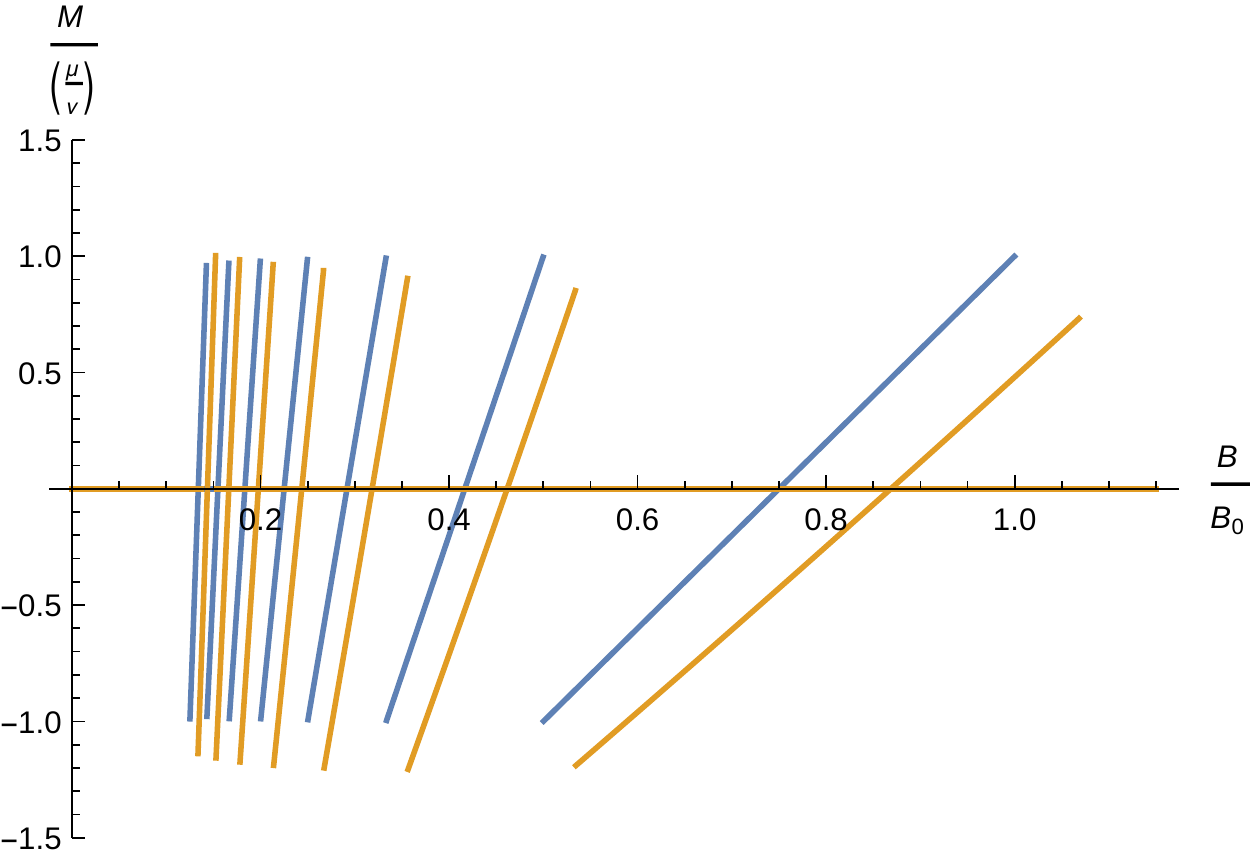}
  \caption{Reduced magnetisation, $\mathcal{M}/(\frac{\mu}{\mathcal{V}})$ vs magnetic field, $y=B/B_0$ plot in de Haas-Van Alphen effect, with the GUP dependent parameter, $\alpha mB_{0} \mu=10^{-2}$. The blue curve indicates without GUP correction,  and the yellow one indicates the result with GUP correction.}
  \label{fig 2}
\end{figure}

\begin{figure}
 \centering
  \includegraphics[scale=0.8]{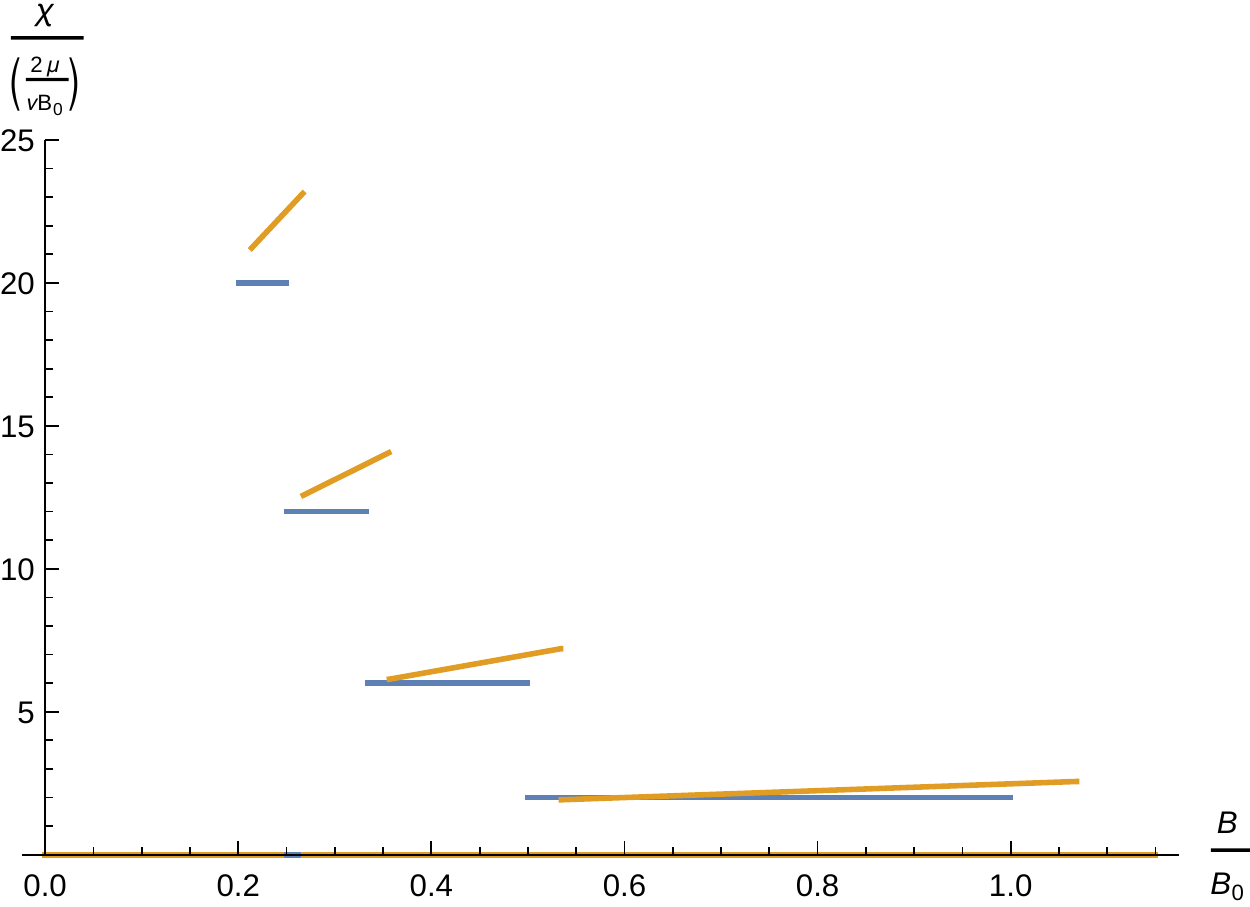}
  \caption{Reduced susceptibility, $\chi/(\frac{2\mu}{\mathcal{V}B_0)}$ vs magnetic field, $y=B/B_0$ plot in de Haas-Van Alphen effect, with the GUP dependent parameter, $\alpha mB_{0} \mu=10^{-2}$. The blue curve indicates without GUP correction,  and the yellow one indicates the result with GUP correction.}
  \label{fig 3}
\end{figure}

We consider $E_{0}$ to be the ground state energy of an ideal Fermi gas at absolute zero, and $E_{0}$ is a function of the field strength $B$. To simplify our calculation, we ignore the motion of the electrons along the direction of magnetic field $B$, i.e. we take $p_{z}=0$. In other words, we consider a two dimensional Fermi gas whose single-particle energy levels are,
\begin{align}
\epsilon_{j} & = \mu B(2j+1)[1 + 4\alpha m\mu B(2j+1)],
\end{align}
 which is $g_{j}$-fold degenerate, with,
\begin{eqnarray}
g_{j} = g[ 1 - 6\alpha m\mu B(2j+1) ],
\end{eqnarray} 
where $g=\frac{V^{2/3}}{2\pi}{eB}$ is constant low energy density of states. The ground state energy $E_{0}$ is the sum of $\epsilon _{j}$ over the lowest $N$ single particle states. Since $j$-th level degeneracy $g_{j}$ depends on the field strength $B$, the maximum number of particles that can have the energy $\epsilon _{j}$ depends on $B$. If the field $B$ is such that $g_{0}\geq N$ then all particles can occupy the lowest energy level $(j=0)$ and $E_{0}$ becomes,
\begin{eqnarray}
E_{0}(B) &=& N \epsilon _{0}=N\mu B[1+4\alpha m \mu B],\nonumber\\
\frac{E_{0}(B)}{N}&=& \mu B_{0} y + 4\alpha m\mu^{2}B_{0}^{2} y^{2};\quad \text{for } g_{0}\geq N,
\end{eqnarray}
where, we have defined two new variables $B_{0}:=\frac{NB}{g} $ and $y:=\frac{B}{B_{0}}$. The variable $B_0$
depends on the system, the volume ($V$), and total number of electrons ($N$). For a particular value of field $B$, suppose $\ \sum_{i=0}^{j} g_{i}\ < N <\ \sum_{i=0}^{j+1} g_{i}\ $, then the $(j+1)$ number of lowest levels are completely filled with $g_{i}$ number of particles in every $i$th level. Also, the $(j+1)$st level is only partially filled, and the higher levels are empty. In this case, the ground state energy in terms of $B_0$ and $y$ becomes,
\begin{align*}
E_{0}(B) &= \sum_{i=0}^{j} g_{i}\epsilon _{i} + \bigg[ N - \sum_{i=0}^{j} g_{i} \bigg] \epsilon _{j+1}\nonumber\\
&=  g\mu B\left[ \sum_{i=0}^{j}(2i+1) - 2\alpha m\mu B\sum_{i=0}^{j}(2i+1)^2 \right]+\mu B\left[N- \left\{ \sum_{i=0}^{j}g - 6\alpha m\mu Bg\sum_{i=0}^{j}(2i+1)\right\} \right]\nonumber\\
&\hspace{8cm}\times\left[ (2j+3) + 4\alpha m\mu B(2j+3)^2 \right]\nonumber\\
&=  g\mu B\left[(j+1)^2 - \frac{2}{3}\alpha m\mu B(j+1)(2j+1)(2j+3) \right]+\mu B\left[N- g\left\{ (j+1) - 6\alpha m\mu B(j+1)^2\right\} \right]\nonumber\\
&\hspace{8cm}\times\left[ (2j+3) + 4\alpha m\mu B(2j+3)^2 \right],
\end{align*}
\begin{align}
\frac{E_{0}(B)}{N} = & \mu B_{0} y [ (2j+3) - y(j+1)(j+2) ] + \frac{2}{3}\alpha m B_{0}^{2}\mu^{2} y^{2}\bigg[ 6(2j+3)^{2} -5y(j+1)(j+2)(2j+3)\bigg];\nonumber\\
&\hspace{10cm}\text{for }\sum_{i=0}^{j} g_{i}\ < N <\ \sum_{i=0}^{j+1} g_{i}.
\end{align}
As calculated in the previous section, the magnetisation of the system is given by, 
\begin{eqnarray}
\mathcal{ M} = -\frac{1}{V}\frac{\partial E_{0}}{\partial B}.
\end{eqnarray}
Now there are two situations when the total number of electrons is greater than the degeneracy of the lowest level with $j=0$, or greater than the degeneracy of any $j$-th level of the system as discussed above. So, the magnetisation,
\begin{eqnarray}
\mathcal{M} &=-\frac{\mu}{\mathcal{V}} \left[ 1 + 8\alpha m B_{0}\mu y \right];\quad \text{for } g_{0}\geq N
\end{eqnarray}
\begin{align}
\mathcal{M}= & \frac{\mu}{\mathcal{V}}\bigg[\bigg( 2(j+1)(j+2)y-(2j+3)\bigg) - 2\alpha mB_{0} \mu  y\bigg(  4(2j+3)^{2}-5y(j+1)(j+2)(2j+3) \bigg)\bigg];\nonumber\\
&\hspace{10cm}\text{for }\sum_{i=0}^{j} g_{i}\ < N <\ \sum_{i=0}^{j+1} g_{i}.
\end{align}
Figure \ref{fig 2}, shows that the GUP corrected magnetisation still changes discontinuously with the magnetic field. Also interestingly the curves approach the low energy results in higher $j$ values.
Now the susceptibility per unit volume is,
\begin{eqnarray}
\chi = \frac{\partial \mathcal{M}}{\partial B} = - \frac{1}{V} \frac{\partial^{2} E_{0}}{\partial B^{2}}.
\end{eqnarray}
Depending upon the electron numbers we have,
\begin{eqnarray}
\chi = -\frac{8\alpha}{\mathcal {V}}m\mu^{2};\quad\text{for }g_{0}\geq N
\end{eqnarray}
\begin{align}
\chi =& \frac{2\mu}{\mathcal{V}B_{0}}(j+1)(j+2) - \frac {2\alpha}{\mathcal{V}}m\mu^{2}\bigg[ 4(2j+3)^{2}-10y (j+1)(j+2)(2j+3) \bigg];\quad\text{for }\sum_{i=0}^{j} g_{i}\ < N <\ \sum_{i=0}^{j+1} g_{i}.
\end{align}
As opposed to the magnetisation case, in Figure \ref{fig 3} the susceptibility approaches to the low energy values in higher $j$ values.
From the above results, it is straightforward to see that the magnetisation $\ (\mathcal{M})\  $ and the magnetic susceptibility per unit volume $(\chi)$ are reduced to the result without minimal length effect as $\alpha\rightarrow 0$. As we have discussed in the previous section, the effect of GUP on this quantum phenomenon is practically undetectable, for the visualisation of results graphically we have taken the value of the GUP parameter to be very large in Figure \ref{fig 2}, and \ref{fig 3}. 
\section{Conclusions}\label{conclusions}
The effect of gravity becomes very interesting at an extremely small length scale, which is very high energy. Thus to address fundamental areas of physics we must resort  to quantum gravity theories. In this article, we have studied  how the theory of Landau diamagnetism gets modified when we consider the effect of quantum gravity theories. Our analysis shows that the overall behavior of the magnetic susceptibility for diamagnetic  material is the same. It gets only a temperature independent constant shift in the first order (first order in $\alpha$) correction due to GUP. In absence of quantum gravity effects, it will reduce to $\frac{1}{T}$ dependence (usual Curie's law). We also investigate the thermodynamical properties of the system at high temperature by calculating the specific heat ($C_V$). GUP effect on $C_V$ is very prominent at extremely high temperature \ref{fig }. At a very low temperature, the susceptibility of an ideal Fermi gas  changes discontinuously with the magnetic field,  commonly known as the de Haas-van Alphen effect. We observed a modified  de Haas-van Alphen effect in presence of GUP as explained in Figure \ref{fig 3}. Although the effect of GUP on our system is very small to detect it experimentally in the lab, we have taken the value of the GUP parameter to be very large to show the corrections graphically. A similar analysis using non-commutative algebra (NC) was done in \cite{Halder:2017lei}, which may help to further investigate the relationship between NC algebra and GUP for this simple system of an ideal Fermi gas.

\section*{Acknowledgements}
MA thanks Aditya Dwivedi and Kajal Singh for useful discussions. BPM acknowledges the research Grant for faculty under IoE Scheme (Number 6031) of Banaras Hindu University, Varanasi.  

{\bf Our manuscript has no associated data}.

\bibliographystyle{unsrt} 
\bibliography{GUP} 
\end{document}